\documentclass[twoside,11pt]{article}

\usepackage{obs_study_style}

\usepackage{url}
\usepackage{tikz}
\usepackage{amsmath}
\usepackage{setspace}
\usepackage{pifont}
\usepackage{amsbsy}
\usepackage{amsfonts}
\usepackage{pgf}
\usepackage{subcaption}

\usetikzlibrary{shapes, decorations, arrows, calc, arrows.meta, fit, positioning, shapes.geometric, shapes.misc, bending}

\tikzset{
  -Latex,semithick,
  >={Latex[width=1.5mm,length=2mm]},
  obs/.style n args = {2}{name = #1, circle, draw, inner sep = 7pt, label = center:$#2$},
  do/.style n args = {2}{name = #1, regular polygon, regular polygon sides = 4, draw, inner sep = 5pt, label = center:$#2$},
  trans/.style n args = {2}{name = #1, regular polygon, regular polygon sides = 3, draw, inner sep = 2pt, label = center:$#2$}
}

\tikzstyle{box} = [rectangle, rounded corners,
text centered,
align=left,
draw=black]

\renewcommand{\indent}{\hspace{5mm}}
\newcommand{\given}{{ \, | \, }}
\newcommand{\doo}{\textrm{do}}

\heading{}{}{}{}{}{Lauri Valkonen, Santtu Tikka, Jouni Helske and Juha Karvanen}

\ShortHeadings{Price Optimization Combining Conjoint Data and Purchase History}{Valkonen, Tikka, Helske, and Karvanen}
\firstpageno{1}

\begin{document}

\def\spacingset#1{\renewcommand{\baselinestretch}%
{#1}\small\normalsize} \spacingset{1}

\title{Price Optimization Combining Conjoint Data and Purchase History: A Causal Modeling Approach}

\author{\name Lauri Valkonen \email lauri.p.k.valkonen@jyu.fi \\
       \addr Department of Mathematics and Statistics \\
    P.O.Box 35 (MaD) FI-40014 University of Jyv\"askyl\"a
       \AND
       \name Santtu Tikka \email santtu.tikka@jyu.fi \\
       \addr Department of Mathematics and Statistics \\
    P.O.Box 35 (MaD) FI-40014 University of Jyv\"askyl\"a
              \AND
       \name Jouni Helske \email jouni.helske@jyu.fi \\
       \addr Department of Mathematics and Statistics \\
    P.O.Box 35 (MaD) FI-40014 University of Jyv\"askyl\"a
              \AND
       \name Juha Karvanen \email juha.t.karvanen@jyu.fi \\
       \addr Department of Mathematics and Statistics \\
    P.O.Box 35 (MaD) FI-40014 University of Jyv\"askyl\"a}

\maketitle

\begin{abstract}
Pricing decisions of companies require an understanding of the causal effect of a price change on the demand. When real-life pricing experiments are infeasible,  data-driven decision-making must be based on alternative data sources such as purchase history (sales data) and conjoint studies where a group of customers is asked to make imaginary purchases in an artificial setup. We present an approach for price optimization that combines population statistics, purchase history, and conjoint data in a systematic way. We build on the recent advances in causal inference to identify and quantify the effect of price on the purchase probability at the customer level. The identification task is a transportability problem whose solution requires a parametric assumption on the differences between the conjoint study and real purchases.  The causal effect is estimated using Bayesian methods that take into account the uncertainty of the data sources. The pricing decision is made by comparing the estimated posterior distributions of gross profit for different prices. The approach is demonstrated with simulated data resembling the features of real-world data.
\end{abstract}

\begin{keywords}
  Bayesian model, Causal inference, Data-fusion, Demand estimation, Pricing, Transportability
\end{keywords}

\section{Introduction}

Pricing decisions are vital for any business seeking profitability. In price optimization \citep{phillips2021pricing}, a company has to estimate the impact of a price change on the behavior of both current customers and potential new customers. This estimation task is essentially a problem of causal inference \citep{pearl2009} where the goal is to quantify the causal effect of price on the demand \citep{guelman2014causal}.

Price elasticity of demand is a traditional approach for measuring the effect of price change in demand \citep{bijmolt2005new}. However, elasticity is only the derivative of demand at one price point and may not correctly describe the change of demand for large price changes. A more general approach is to consider the price-response function  \citep{phillips2021pricing} that specifies the expected demand as a function of price. When customer-level data are available, the purchase probabilities of each customer can be modeled as a function of price and customer characteristics.

Historical data on sales and prices are readily available for most companies. If the product is a subscription-based digital service, purchases can be identified on the customer level and additional personal data on demographics and usage habits can be often collected via registration forms and browser cookies. Historical sales data alone are usually insufficient for price optimization. The price may have remained the same for a relatively long time and there may be confounders that have affected both the changes in pricing and the changes in sales \citep{tian2020optimizing}.

In a pricing experiment, also known as an A/B test \citep{kohavi2017online}, customers are randomized into two or more groups and a different price is offered for each group. Ideally, a pricing experiment is the most reliable way to estimate the causal effect of price on the purchase. However, some challenges may render real-life pricing experiments impractical. For instance, browser cookies may not have one-to-one correspondence with customers, which complicates the technical implementation of randomization in digital channels. In addition to the lack of full controllability, offering different prices may cause confusion among customers, affect their behavior, and even have negative effects on the brand.

A conjoint study \citep[see e.g.,][]{rao2014applied} is an experimental setting, where a customer is asked to choose among different variations of a product with modifications to the features, such as price. A conjoint study as a well-designed randomized experiment can bring valuable information about the customers' sensitivity to prices and enable testing of many different prices at once. However, the behavior of the participants in the artificial setup may differ from their behavior in a real purchase situation \citep[see e.g.,][]{natter2002real}.

Combining data collected under various potentially heterogeneous conditions is known as the data-fusion problem \citep{bareinboim_data_fusion}. Such heterogeneity can be a result of inherent differences between the populations or the sampling methods used, for example. Data-fusion is often not straightforward, as in the case of combining data from pricing experiments and conjoint studies due to the difference in participant behavior. This scenario is an example of a transportability problem where the goal is to use data from several source populations to make inferences about a target population of interest.

In this paper, we propose an approach for combining different data sources to estimate the causal effect of pricing on a purchase in the absence of real-life pricing experiments. We consider a subscription-based service, such as video or music streaming, an audiobook service, or a digital newspaper. The proposed work is motivated by a real business case that we carried out together with a company that offers subscription-based digital services. As publishing the results based on the real data would be against the business interest of the company, we use simulated data that aim to capture the most important features of real data.

The proposed approach consists of four steps: 1) The causal relations of the purchase process are described in the form of a directed acyclic graph (DAG). 2) The causal effect of price on purchases is identified from both observational and experimental data sources presented in a symbolic form. 3) A hierarchical Bayesian model is fitted to estimate the causal effect according to the obtained identifying functionals. 4) The posterior distribution of the optimal price is found by maximizing the expected gross profit defined as a function of the price and the purchase probabilities estimated in step 3.

The rest of the paper is organized as follows. We begin by describing the subscription-based business model, the data sources, and the transportability problem in Section~\ref{sec:problem}. The details of the simulation are described in Section~\ref{sec:simulation}. The optimization procedure and its results are presented in Section~\ref{sec:optimization} and compared with the approach where only sales data are used for estimation and optimization. We conclude the paper with an evaluation of our method and discuss further research possibilities in Section~\ref{sec:discussion}. Table~\ref{tab:notation} in the Appendix summarizes the notation of this paper. The code and the data for reproducing the results and the figures of the paper are available at \url{https://github.com/lvalkone/Proptim}.

\section{Causal model for a subscription-based service}\label{sec:problem}

\subsection{Problem definition and causal diagram}
We consider a company that offers a subscription-based service for consumers. The subscription is automatically renewed monthly unless the customer cancels it. The total revenue of the company is the product of the price and the number of subscriptions, and the total profit is the difference between the revenue and the costs which can be divided into variable costs and fixed costs. The total profit is maximized when the gross profit, defined as the difference between the revenue and the variable costs, is maximized.

In addition to the price, the purchase behavior of a customer depends on background variables such as age, gender, and location. It is also clear that the probability of retention (an existing customer continues to subscribe) is higher than the probability of acquisition (a new customer subscribes) \citep{reinartz2005balancing}. In our model, we assume that each customer has a latent personal reference price \citep[see e.g.,][]{refprice_winer_russel, mazumdar2005reference, CAO2019540} that they implicitly or explicitly compare with the actual price. When the price of the product is changed, the subscription probability will change as well, and the impact will be different for new and existing subscriptions because the distribution of reference prices differs between these groups due to selection. Estimating the demand after the price change is essential for the optimization of the profit.

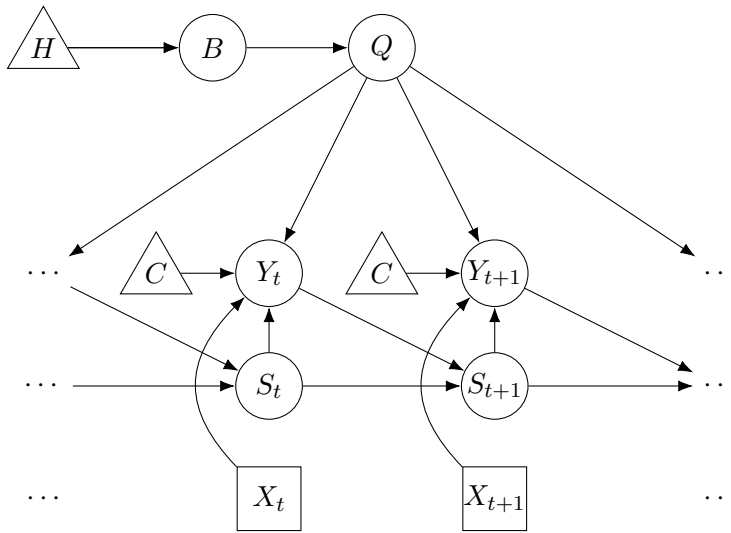
\begin{figure}
\centering
\begin{tikzpicture}[xscale=3,yscale=1.5]
\node [trans = {h}{H}] at (-1,2) {$\vphantom{0}$};
\node [obs = {b}{B}] at (-0.25,2) {$\vphantom{0}$};
\node [obs = {q}{Q}] at (0.5,2) {$\vphantom{0}$};
\node [trans = {c1}{C}] at (-0.5,0) {$\vphantom{0}$};
\node [trans = {c2}{C}] at (0.5,0) {$\vphantom{0}$};
\node (ypast) at (-1,0) {${\cdots}$};
\node [obs = {y1}{Y_{t}}] at (0,0) {$\vphantom{0}$};
\node [obs = {y2}{Y_{t+1}}] at (1,0) {$\vphantom{0}$};
\node (yfuture) at (2,0) {${\cdots}$};
\node (spast) at (-1,-1) {$\cdots$};
\node [obs = {s1}{S_{t}}] at (0,-1) {$\vphantom{0}$};
\node [obs = {s2}{S_{t+1}}] at (1,-1) {$\vphantom{0}$};
\node (sfuture) at (2,-1) {${\cdots}$};
\node (xpast) at (-1,-2) {${\cdots}$};
\node [do = {x1}{X_{t}}] at (0,-2) {$\vphantom{0}$};
\node [do = {x2}{X_{t+1}}] at (1,-2) {$\vphantom{0}$};
\node (xfuture) at (2,-2) {${\cdots}$};
\draw [->] (h) -- (b);
\draw [->] (c1) -- (y1);
\draw [->] (c2) -- (y2);
\draw [->] (h) -- (b);
\draw [->] (b) -- (q);
\draw [->] (q) -- (ypast);
\draw [->] (q) -- (yfuture);
\draw [->] (q) -- (y1);
\draw [->] (q) -- (y2);
\draw [->] (ypast) -- (s1);
\draw [->] (spast) -- (s1);
\draw [->] (y1) -- (s2);
\draw [->] (s1) -- (y1);
\draw [->] (s1) -- (s2);
\draw [->] (s2) -- (y2);
\draw [->] (y2) -- (sfuture);
\draw [->] (s2) -- (sfuture);
\draw [->] (x1) to [bend left=27] (y1);
\draw [->] (x2) to [bend left=27] (y2);
\end{tikzpicture}
\caption{Cross-section of the data-generating process for the purchase decisions at time points $t$ and $t+1$. Circles denote observed variables, triangles denote transportability nodes, and squares denote variables targeted by interventions. \label{dag}}
\end{figure}

Figure~\ref{dag} shows the DAG that represents the causal relations of the variables of interest in our scenario. At the time $t$, the purchase $Y_t$ is affected directly by the price of the service $X_t$. We also assume that the purchase $Y_t$ is affected by a latent variable $Q$ that represents the customer's reference price. The larger the difference between the reference price and the product price is, the more likely the customer is willing to buy. The reference price is also affected by the background variables (age, gender, location) denoted commonly by cluster~$B$ \citep{tikka2021clustering}, which for simplicity of the exposition are assumed to be constant in time. Accumulated subscription history is assumed to have a positive effect on repurchasing, and thus we also assume that $Y_t$ is affected by the number of consequent subscription periods $S_t$.

The DAG also contains transportability nodes that describe differences between populations in the underlying causal model \citep{bareinboim_transportability}. The transportability node $H$ describes that the distribution of the background variables $B$ may differ between subscribers and non-subscribers. The transportability node $C$ (presented separately for time points $t$ and $t+1$) describes that the purchase probability of a customer may be different between a conjoint scenario and a real purchase scenario.

\subsection{Data sources}\label{sec:datasources}
The company gathers data on the customers, purchases $Y_t$, and the purchase history $S_t$. The customers register for the service and provide information on their age ($A$), gender ($G$), and location ($L$), commonly denoted by $B$. We denote the distributions related to the subscriber population ($H = 1$) as $P^{(1)}$ and to the non-subscriber population ($H = 0$) as $P^{(0)}$.

The company conducts a conjoint study \citep{rao2014applied} to increase its understanding of the expected consequences of a price change.
This allows the company to test a variety of prices without interfering with the actual business. For simplicity, we do not explicitly consider different product features in the conjoint study but we point out that they could be studied as well.
The conjoint is targeted at both current and earlier subscribers as well as those who have never subscribed.

In a symbolic form, the purchase history data is denoted by $P^{(1)}(Y_t, S_t \given \doo(X_t), Q)$. The conjoint data contains both subscribers and non-subscribers and thus provides information from two domains via two distributions: $P^{(1)}(Y_t, S_t \given \doo(X_t), Q, C)$ and $P^{(0)}(Y_t, S_t \given \doo(X_t), Q, C)$, respectively. The purchase history and the conjoint data cannot be combined straightforwardly because the behavior of customers is expected to differ between the real purchase situation and the artificial purchase situation in the conjoint study. Formally, we can show that our causal effects of interest, i.e., the effect of price on the purchase in the context of a real purchase scenario for subscribers and non-subscribers, respectively, cannot be identified from the conjoint data alone because the transportability node $C$ directly affects the purchase decision $Y$. Therefore, the conjoint data cannot be used without further assumptions. We assume here that the difference in the purchase probability between the conjoint scenario and the real purchase scenario can be parametrized via a level shift in the reference price:
\begin{equation} \label{eq:conjointeffect}
  P^{(i)}(Y_t, S_t \given \doo(X_t), Q = q) = P^{(i)}(Y_t, S_t \given \doo(X_t), Q = q + \kappa, C), \quad i = 0, 1,
\end{equation}
where the unknown parameter $\kappa$ models the conjoint effect. In reality, the difference between the scenarios could be more complicated and depend on other factors as well, but here we chose the level shift for simplicity. In general, the more complicated the parameterization for the difference between the scenarios is, the more data will be required to estimate the difference. The impact of model misspecification is studied in Section~\ref{sec:misspecification}.

Population-level data on the background variables are needed when potential new customers are included in the analysis. These data are available in an aggregated form from official statistics, and they provide information on the distribution $P^{(0)}(B)$. Data on the background variables of the subscribers is available from the price history data as $P^{(1)}(B)$. We assume that the reference price is a latent variable and that the parametric form of its distribution $P^{(0)}(Q|B)$ is known. Given the background variables, the distribution of the reference price is the same across domains, i.e., $P^{(0)}(Q|B) = P^{(1)}(Q|B)$, which we simply denote by $P(Q|B)$.

\subsection{Identifying the causal effect and formulating the model}\label{identification}

Our goal is to identify the causal effect of price on the purchase probability for non-subscribers and subscribers, i.e., $P^{(0)}(Y_t \given \doo(X_t))$ and $P^{(1)}(Y_t \given \doo(X_t))$ from the available data sources in the DAG of Figure~\ref{dag}. Due to the challenging nature of the task, we apply the do-search algorithm \citep{dosearch2021} to identify the effects. Do-search accepts arbitrary data sources (in a symbolic form) as input and provides the identifying functional when the effect in question is identifiable. The DAG of Figure~\ref{dag} and the data sources $P^{(0)}(Y_t, S_t \given \doo(X_t), Q)$, $P^{(1)}(Y_t, S_t \given \doo(X_t), Q)$, $P^{(0)}(B)$, $P^{(1)}(B)$, and $P(Q \given B)$ are given as input to the algorithm. We note that the conjoint data also provides information on $P^{(0)}(Y_t, S_t \given \doo(X_t), Q)$ and $P^{(1)}(Y_t, S_t \given \doo(X_t), Q)$ due to our assumption in Equation~\eqref{eq:conjointeffect}. Do-search returns the following identifying functionals for our causal effects of interest:
\begin{align*}
  P^{(0)}(Y_t \given \doo(X_t)) &= \sum_{B, S_t}\int_{Q} P^{(0)}(Y_t \given \doo(X_t),Q,S_t) P^{(0)}(S_t \given \doo(X_t), Q) P(Q|B) P^{(0)}(B), \\
  P^{(1)}(Y_t \given \doo(X_t)) &= \sum_{B, S_t}\int_{Q} P^{(1)}(Y_t \given \doo(X_t),Q,S_t) P^{(1)}(S_t \given \doo(X_t), Q) P(Q|B) P^{(1)}(B).
\end{align*}
We need to model the conditional distribution $P(Q|B)$ of the latent reference price $Q$. As the reference price is assumed to vary between individuals, we fit a log-normal model for each individual $i$. Our model for the reference price is then
\begin{equation} \label{eq:Q_def}
  Q_i = \exp(\beta_0 +  \beta_{1, A_i} + \beta_{2, G_i} + \beta_{3,L_i} + u_i),
\end{equation}
where $\beta_{j,k}$ refers to the $k$th category of the $j$th predictor, and $u_i$ is an individual-specific normally distributed random effect. The purchase choice is modeled via a Bernoulli distribution using the logit-link as
\begin{equation} \label{eq:purchase_logit_def}
  \mathrm{logit}(\pi_{i,t}) = \alpha_1(Q_i + I(C_i = 1)\kappa - X_{i,t}) + \alpha_2 \log(S_{i,t} + 1) + \alpha_3 I(S_{i,t} = 0),
\end{equation}
where the parameter $\alpha_1$ describes the impact of the difference between the reference price $Q_i$ (adjusted by the conjoint effect $\kappa$ when $C_i=1$) and the price $X_{i,t}$, the parameter $\alpha_2$ is the effect of consecutive subscription periods, and the parameter $\alpha_3$ describes the impact of the difference between customers who are just starting their subscription and those who are simply continuing their earlier subscription.

\section{Data simulation} \label{sec:simulation}

We simulate our data based on the structure of the DAG in Figure~\ref{dag}. The simulations are implemented in the R environment \citep{rsoftware} with the R6causal package \citep{r6causal}. The statistics related to model parameters are described in Table~\ref{param_estimates} of the Appendix.

We use real data from Statistics Finland \citep{StatFin} as a basis for the population demographics, covering a joint distribution of age, gender, and location. The variables are categorized for our study such that the age covers four groups: 18--30, 31--45, 46--60, and 61--75. The two other variables consist of dichotomous classifications as male and female in gender, and urban and rural in location. According to the data above, the company's target market is limited to Finland and individuals who are between 18--75 years old. The market size of the population of our study in 2020 is thus $N = 3\,956\,294$. Other variables of the DAG are simulated according to the flowchart presented in Figure~\ref{simulation_flow}. The chosen parameters defining the functional forms of the causal relationships used in the simulation are presented in Table \ref{param_estimates} of the Appendix.

At the time $t=1$, the company launches the product to the market, and every time point~$t$, a share of the population signs into the service of which a subset makes a subscription. We simulate the subscription history of two years ($t=1,2,\ldots,24$). The company launches the product for 16 euros. The price is then raised by 0.50 euros two times during the two years: after 6 months and after 18 months, i.e., at the times $t=7$, and $t=19$.

The customer choice (purchase or not) is modeled by a Bernoulli distribution with the purchase probability $\pi_{i,t}$. The probability $\pi_{i,t}$ depends on the difference between the product price and the customer's reference price which is affected by the background variables $B$ (see the DAG in Figure~\ref{dag}). Besides this, the number of earlier subscription periods $S_{i,t}$ directly affects the choice.

\newpage
\begin{figure}[ht!!]
\centering
\begin{tikzpicture}
\spacingset{1}
\centering
\scriptsize
\node (init) [box, label=above:\textbf{1. Initialize:}] at (0, 0){\\
$\bullet$ Size of the population $ \mathcal{D}^{pop}$: $N=3956294 $\\
$\bullet$ Price of the product at time $t=1$: $X_1=16$\\
$\bullet$ Number of current customers at time $t=1$: $n=0$\\
$\bullet$ Number of potential customers at time $t=1$: $n_0=1000$\\
$\bullet$ Number of customers who purchased at time $t=1$: $n_1=0$\\
$\bullet$ Potential new and current customers at time $t=1$: $\mathcal{D}^0 = \emptyset$ and $\mathcal{D}^1 = \emptyset$\\
$\bullet$ Number of consecutive subscription periods for all $i=1,\ldots,N$ at time $t=1$: $S_{i,1}=0$,\\
$\bullet$ Parameter values (see Table \ref{param_estimates} in the Appendix): $\beta_0, \beta_{1,A}, \beta_{2,G}, \beta_{3,L}, \kappa, \alpha_1, \alpha_2, \alpha_3, \tau$\\
$\bullet$ Generate individual deviations in the reference prices: $u_i \sim \mathrm{N}(0, \tau^2)$, $i=1,\ldots,N$};

\node (simpur) [box, label=above:\textbf{2. Simulate purchase history for $t=1,\ldots,24$:}, below of = init] at (0, -4.5){\\
$\bullet$ Set price:\\
\indent if ($t=7$ or $t=19$) $X_t = X_{t-1} + 0.5$\\
\indent if ($t=25$) $X_t = x$\\
\indent else $X_t = X_{t-1}$\\
$\bullet$ Set $\mathcal{D}^0$:\\
\indent Sample $n_0$ new customers randomly from $\mathcal{D}^{pop} \setminus \mathcal{D}^1$ \\
$\bullet$ Set $\mathcal{D}$ and update $n$:\\
\indent $\mathcal{D} = \mathcal{D}^0 \cup \mathcal{D}^1$\\
\indent$n=n_0 + n_1$\\
$\bullet$ Calculate reference price and choice probability for all $i=1,\ldots,n$ in $\mathcal{D}$:\\
\indent $Q_i = \exp(\beta_0 +  \beta_{1, A_i} + \beta_{2, G_i} + \beta_{3,L_i} + u_i)$\\
\indent $\mathrm{logit}(\pi_{i,t}) = \alpha_1(Q_i - X_{i,t}) + \alpha_2 \log(S_{i,t} + 1) + \alpha_3 I(S_{i,t}=0)$\\
$\bullet$ Generate customer choices for all $i=1,\ldots,n$ in $\mathcal{D}$:\\
\indent $Y_{i,t} \sim \mathrm{Ber}(\pi_{i,t})$\\
$\bullet$ Update customer data: \\
\indent $S_{i,t+1} = I(Y_{i,t} = 1)(S_{i,t} + 1)$\\
\indent $n_1 = \sum({\mathcal{D} \given Y=1})$\\
\indent $\mathcal{D}^1 = (\mathcal{D} \given Y=1)$
};

\node (simcon) [box, label=above:\textbf{3. Simulate conjoint studies:}] at (0, -12.4) {
\\
$\bullet$ Initialize:\\
\indent $\circ$ Conjoint effect: $\kappa = 0.75$ \\
\indent $\circ$ Number of participants in each conjoint study: $m=200$\\
\indent $\circ$ Number of tasks shown per participant: $k=10$\\
\indent $\circ$ Price alternatives: $\mathbf{x}^C=12,12.5,\ldots,21.5,22$\\
$\bullet$ Define the conjoint populations:\\
\indent $\circ$ $\mathcal{C}^0$: Sample $m$ customers randomly from $\mathcal{D}^0$\\
\indent $\circ$ $\mathcal{C}^1$: Sample $m$ customers randomly from $\mathcal{D}^1$\\
\indent $\circ$ $\mathcal{C}^2$: Sample $m$ customers randomly from $\mathcal{D}^{pop} \setminus (\mathcal{D}^0 \cup \mathcal{D}^1)$\\
$\bullet$ For all participants $i$ in $\mathcal{C}^0$, $\mathcal{C}^1$, and $\mathcal{C}^2$:\\
\indent $\circ$ Create price combinations $\mathbf{X}^{C}_i$ by taking $k$ random samples without replacement from $\mathbf{x}^C$\\
\indent $\circ$ For all $\mathbf{X}^{C}_i$\\
\indent\indent Calculate reference price and choice probability:\\
\indent\indent\indent $Q_{i} = \exp(\beta_0 +  \beta_{1, A_i} + \beta_{2, G_i} + \beta_{3,L_i} + u_i)$\\
\indent\indent\indent $\mathrm{logit}(\boldsymbol{\pi}_i) = \alpha_1(Q_i + \kappa -\mathbf{X}^{C}_{i}) + \alpha_2 \log(S_{i} + 1) + \alpha_3 I(S_{i}=0)$\\
\indent\indent Generate customer choices \\
\indent\indent\indent $\boldsymbol{Y}_i \sim \mathrm{Ber}(\boldsymbol{\pi}_i)$};

\node (simnew) [box, label=above:\textbf{4. Simulate purchase history for $t=25$:}] at (0, -16.4) {\\
$\bullet$ Simulate purchase histories with each price $x$ from 14 to 18 euros by 0.25 euros, using block 2.
};
\normalsize
\end{tikzpicture}
\caption{Algorithmic description of the data simulation: The first block describes the elements that have to be initialized for the purchase history simulation in the second block. The third block describes the simulation of the conjoint studies after $t=24$, and the fourth block returns to the second block for future purchase simulations of $t=25$.}
\label{simulation_flow}
\end{figure}
\newpage

Using the sales data alone to find the optimal price is unreliable because it contains information on the effect of the price for only three points: 16, 16.5, and 17 euros. Thus, after two years in the market, the company implements a conjoint study to explore the price sensitivity of customers. The participants of conjoint studies include current and earlier customers as well as those who have never subscribed.

In the conjoint study setting, 10 separate tasks of product alternatives are shown to a customer who is asked to choose whether to purchase the option offered. The alternatives of the product consist of varying prices ranging between 12 and 22 euros by 0.50 euros. With this price range, 10 out of 21 possible tasks are randomly sampled without replication for a participant to be shown. The conjoint effect $\kappa$ of Equation~\ref{eq:conjointeffect} is set to a moderate value of 0.75 euros.

\section{Optimizing the price} \label{sec:optimization}

\subsection{Estimating the model}

We estimate a joint Bayesian \citep[see e.g.,][]{gelmanbayes} model for the reference prices and purchases using Markov chain Monte Carlo via the NIMBLE software and the R package nimble \citep{nimble}. Specifically, we use dynamic Hamiltonian Monte Carlo and the No-U-Turn Sampler \citep{Hoffman2014} provided by the nimbleHMC package \citep{nimbleHMC}. Prior distributions for the regression coefficients are defined as $N(0,0.5)$ distributions. Individual random effects $u_i$ are assumed to be distributed as $N(0, \tau^2)$, and for the standard deviation $\tau$ we set a prior $\tau \sim \mbox{Gamma}(2, 0.2)$ using the shape-scale parameterization.

Table~\ref{param_estimates} shows the posterior statistics and convergence diagnostics for the non-individual-specific parameters. The model has converged well according to these statistics: $\hat{R}$ values are less than 1.01 and the effective sample sizes are sufficient for all parameters, including those not shown in Table~\ref{param_estimates} \citep[see][for details on these diagnostics]{Vehtari2021}.

\subsection{Maximizing the expected gross profit} \label{sec:maximization}

Next, we describe how to optimize the price of the product by using the Bayesian model defined in Section~\ref{identification}. We aim to maximize the expected gross profit at the time $t$, defined as
\begin{equation}\label{totprof}
f(x, \mu^{0}(x,t), \mu^{1}(x,t) \,| N^0(t), N^1(t), \mathcal{V}) = N^0(t)\mu^0(x,t)(x-\mathcal{V}) + N^1(t)\mu^1(x,t)(x-\mathcal{V}),
\end{equation}
where $x$ is the price of the product, $\mu^0(x,t)$ and $\mu^1(x,t)$ are the purchase probabilities of earlier and current customers, respectively. $N^0(t)$ is the number of potential customers making a purchase decision at time $t$, whereas $N^1(t)$ is the number of current customers i.e., those who purchased at time $t-1$ and are now deciding to continue or cancel the subscription. The cost structure related to the gross profit includes the variable costs, which are denoted as $\mathcal{V}$.

As we optimize the expected gross profit by the price at the following time point $t=25$, we can formalize the optimization problem as
\begin{align*}
&\max_{x} \left( f(x, \mu^0(x,t), \mu^1(x,t) \,|\, N^{0}(25), N^{1}(25), \mathcal{V}) \right) \\
= &\max_{x} \left( N^0(25)\mu^0(x,25)(x-\mathcal{V}) + N^1(25)\mu^1(x,25)(x-\mathcal{V}) \right).
\end{align*}
We assume that every month a constant number of potential customers $N^0(t)=1000$ make a purchase choice. On the other hand, the number of current customers $N^1(t)$ can be calculated directly from the historical sales data, which is $1110$ for $t=25$. In addition, we assume that the variable costs $\mathcal{V}$ are constant over time, such that $\mathcal{V}=5$.

In Equation~\eqref{totprof}, the pricing affects both current and potential customers. Therefore, we divide the upcoming customer base data $\mathcal{D}$ into two sets: The data set $\mathcal{D}^0$ includes the customers that either canceled the subscription at $t \leq 24$ or have never purchased. This data set of customers can be obtained by random sampling from the population with a size of $N^0(25)=1000$. The data set $\mathcal{D}^1$ consists of those customers who purchased at $t=24$, so they are automatically about to choose at time $t=25$. This customer base in $\mathcal{D}^1$ can be obtained directly from the historical sales data.

We simulate the posterior distributions of the model parameters using 12 parallel chains, each with 20\,000 iterations, a burn-in period of 2\,000, and a thinning interval of 50 resulting in the final sample size of 4\,800. We denote the model parameters commonly as $\mathbf{\Theta}$. To study the effect of pricing by inspecting the posterior predictive distributions, we repeat the following for a set of different prices: For a given posterior sample, we calculate the reference prices $Q^{\mathcal{D}^0}_i$ and $Q^{\mathcal{D}^1}_j$ and purchase probabilities $\pi^{\mathcal{D}^0}_i$ and $\pi^{\mathcal{D}^1}_j$ for all $i = 1, \ldots, N^0(25)$ and $j = 1, \ldots, N^1(25)$. As none of the customers in $\mathcal{D}^1$ have an estimated individual level $u_j$, we generate the value from a $N(0, \tau^2)$-distribution. The expected purchase probabilities for decisions $Y^0$ and $Y^1$ for the $k$th posterior sample $\mathbf{\Theta}^{k}$ are then estimated as
\begin{align}
\mu^{0,k} &= E(Y^0 \given \doo(X), C = 0, \mathbf{\Theta}^k) \approx \frac{1}{N^0(25)} \sum_{i=1}^{N^0(25)} \pi^{\mathcal{D}^0, k}_i,\label{mu0} \\
\mu^{1,k} &= E(Y^1 \given \doo(X), C = 0, \mathbf{\Theta}^k) \approx \frac{1}{N^1(25)} \sum_{j=1}^{N^1(25)} \pi^{\mathcal{D}^1, k}_j.\label{mu1}
\end{align}
Using the estimates from \eqref{mu0} and \eqref{mu1}, we compute the expected gross profit \eqref{totprof} for each posterior sample $\mathbf{\Theta}^{k}$ to obtain a posterior sample of gross profit values. Finally, the mean and 95\% quantiles are calculated over these posterior samples to obtain the estimated expected gross profits and their credible intervals.

\subsection{Optimization results}

\begin{figure}[!htb]
\centering
\includegraphics[width=\textwidth]{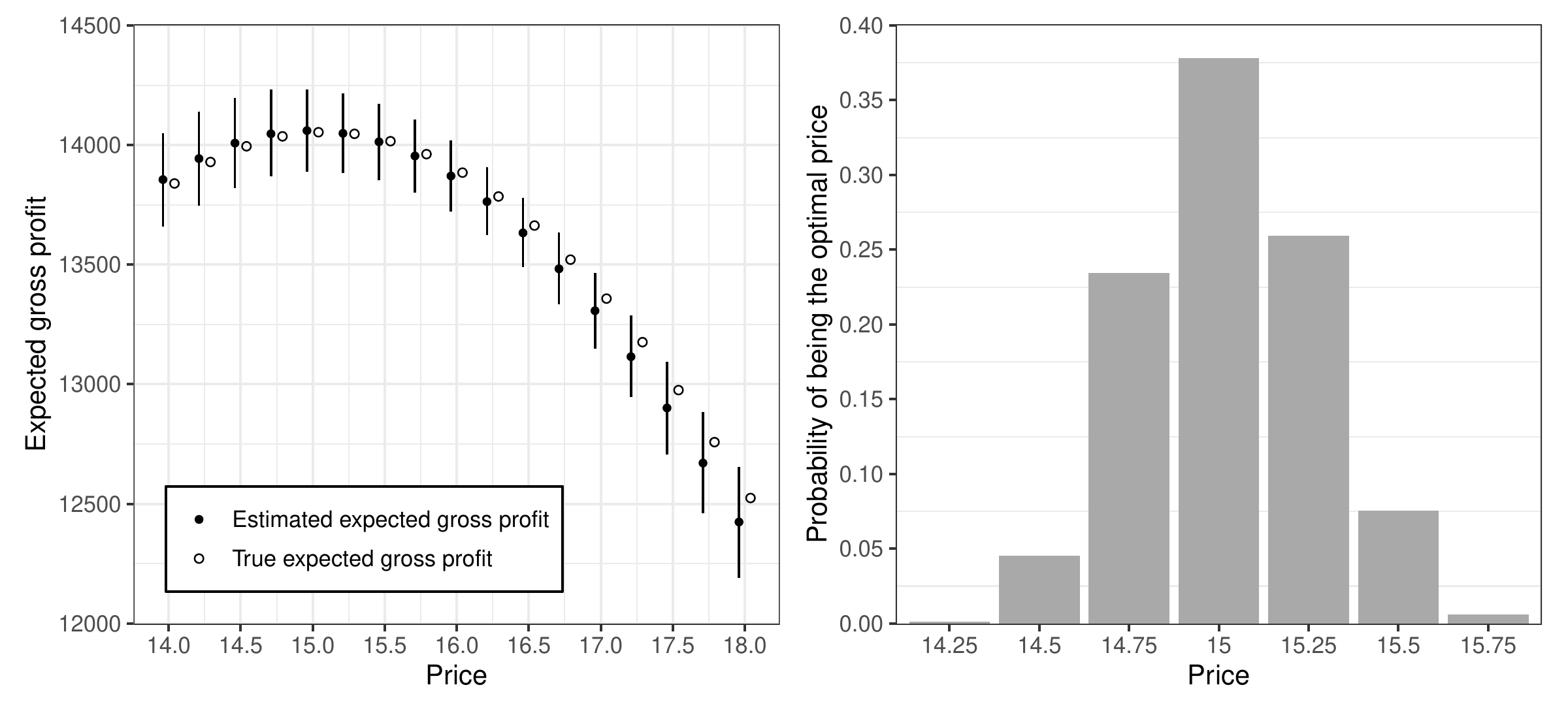}
\caption{Price optimization for time $t=25$. In the left panel, the estimated mean expected gross profits with 95\% credible intervals for different price interventions and the true expected gross profits are presented. The right panel depicts the probability of each price being the optimal price.}
\label{result_figure}
\end{figure}

Figure~\ref{result_figure} shows the results of the price optimization in the case of combining the purchase history and conjoint data. The left panel in Figure~\ref{result_figure} shows the estimated expected gross profits and their 95\% credible intervals for the different price interventions (from 14 to 18 euros by increments of 0.25 euros), and the true expected gross profits, which can be calculated from \eqref{totprof}, where $\mu^0$ and $\mu^1$ are obtained from the purchase history simulation for $t=25$. By calculating which price yields the highest expected gross profit in each posterior sample, we can also obtain a probability distribution of the prices that maximize the expected gross profit, which is presented in the right panel in Figure~\ref{result_figure}. The price with the highest probability, i.e., the optimal price, can be found at $X=15$ with the mean expected gross profit of $14\,060$ euros and $(13\,887, 14\,231)$ 95\% credible interval.

We also attempted to estimate the model and optimize the price without the conjoint data. As expected, we encountered severe convergence issues. This may indicate that the model is not identifiable from the purchase history data alone when the number of different prices is small.

\subsection{Model misspecification} \label{sec:misspecification}

\begin{figure}
  \centering
  \begin{subfigure}{\textwidth}
    \centering
    \includegraphics[width=0.8\textwidth]{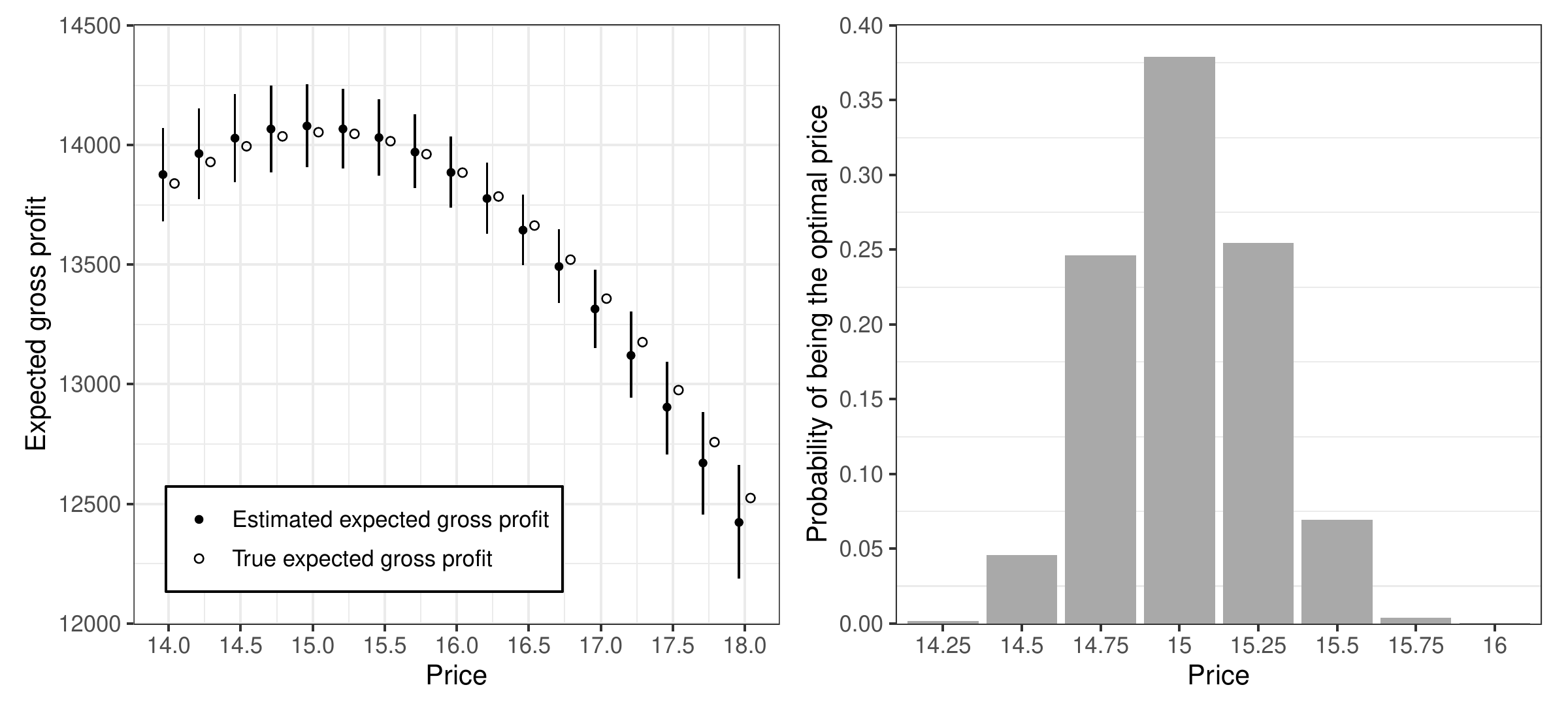}
    \caption{Price optimization for the misspecified reference price $Q_i$.}
    \label{fig:opt_scenario1}
  \end{subfigure}
  \begin{subfigure}{\textwidth}
    \centering
    \includegraphics[width=0.8\textwidth]{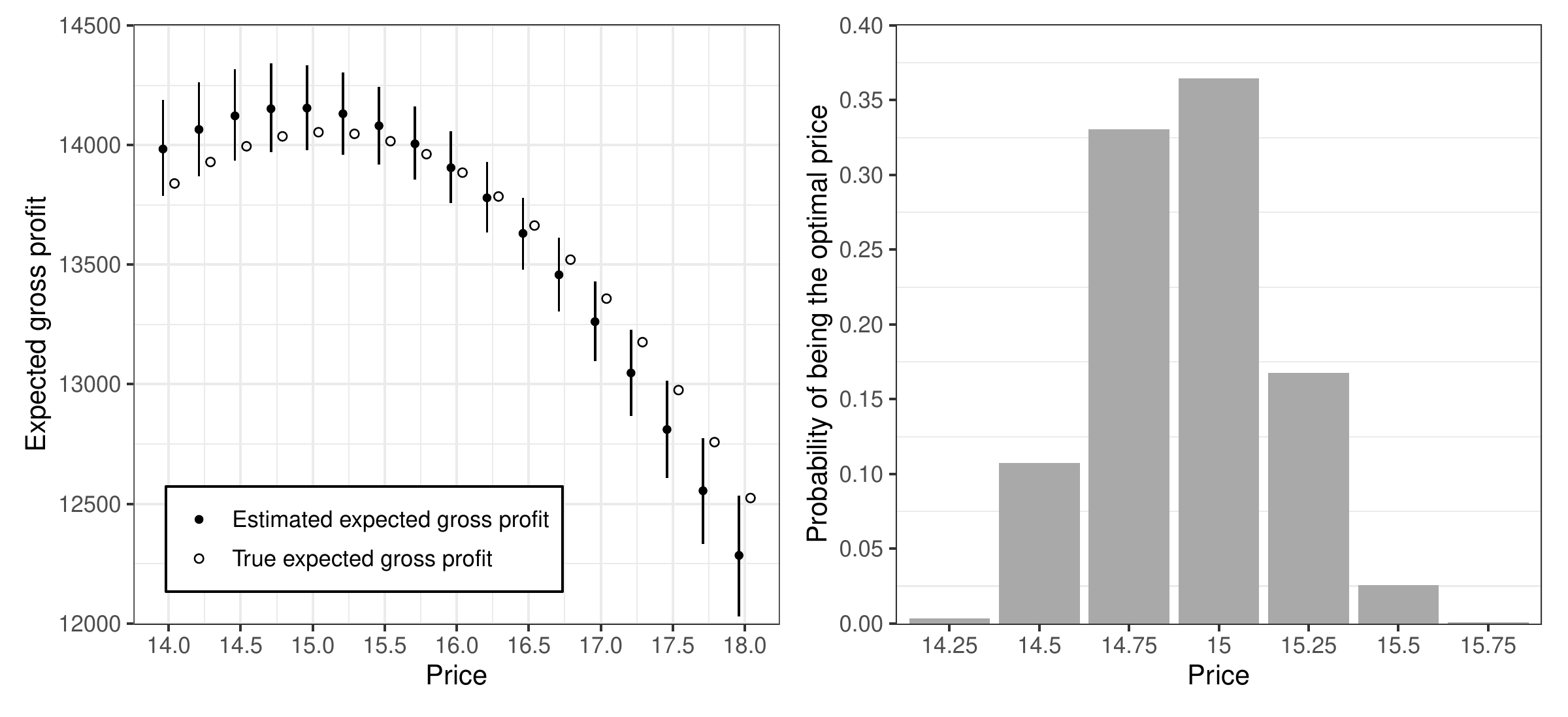}
    \caption{Price optimization for the misspecified conjoint effect $\kappa$.}
    \label{fig:opt_scenario2}
  \end{subfigure}
  \begin{subfigure}{\textwidth}
    \centering
    \includegraphics[width=0.8\textwidth]{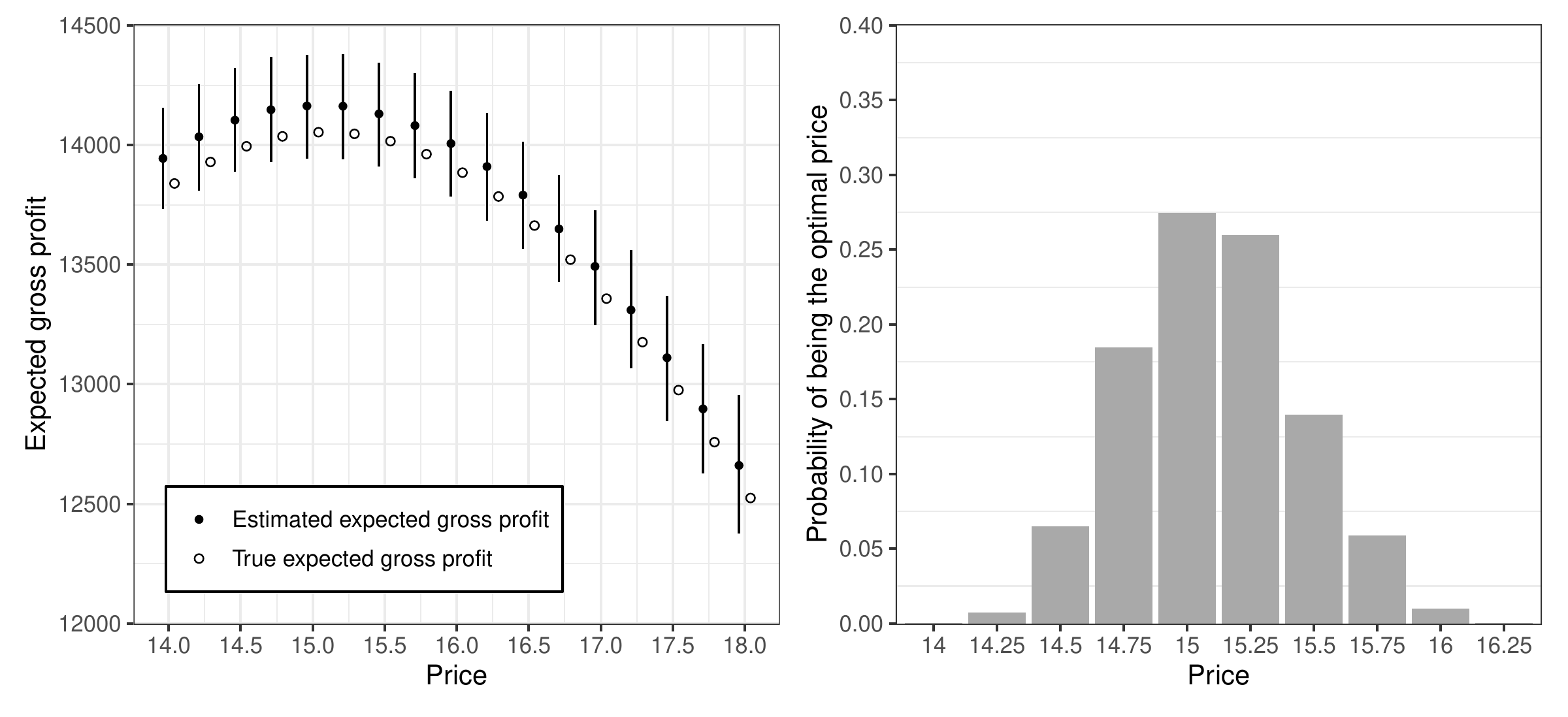}
    \caption{Price optimization for the misspecified effects of the consecutive subscription periods $S_{i,t}$.}
    \label{fig:opt_scenario3}
  \end{subfigure}
  \caption{Optimization results for the misspecified models for time $t=25$. In the left panels, the estimated mean expected gross profits with 95\% credible intervals for different price interventions and the true expected gross profits are presented. The right panels depict the probabilities of each price being the optimal price.}
  \label{fig:opt_scenarios}
\end{figure}

As a robustness check, we evaluate the impact of model misspecification to the prize optimization using the same purchase history and conjoint data as in in Section~\ref{sec:maximization}. Instead of estimating the model based on Equations~\eqref{eq:Q_def} and \eqref{eq:purchase_logit_def}, we consider three scenarios:
\begin{enumerate}
  \item Misspecified reference price $Q_i$ in Equation~\eqref{eq:Q_def}.
  \item Misspecified conjoint effect $\kappa$ in Equation~\eqref{eq:purchase_logit_def}.
  \item Misspecified effects of the consecutive subscription periods $S_{i,t}$ in Equation~\eqref{eq:purchase_logit_def}.
\end{enumerate}
In the first scenario, we omit the characteristics of the individual, i.e., age, gender, and location, from the specification of $Q_i$ in Equation~\eqref{eq:Q_def} leaving only the baseline $\beta_0$ and the individual-specific effect $u_i$ while keeping the model otherwise the same as before. In the second scenario, we incorrectly assume that the effect of $\kappa$ is multiplicative instead of additive, meaning that the predictor related to $\alpha_1$ is now $(Q_i - \kappa^{I(C_i = 1)}X_{i,t})$, but otherwise the model is unchanged. In the third scenario, we ignore the effects of previous consecutive subscription periods from the purchase choice model, using only the effect of the price and its relation to the reference price. In all scenarios, we repeat the prize optimization outlined in Section~\ref{sec:maximization} and the results are shown in Figure~\ref{fig:opt_scenarios}.

The results for the first scenario shown in Figure~\ref{fig:opt_scenario1} indicate that the differences to the estimates obtained from the true model are negligible for this model, which is to be expected as the effects of the demographical characteristics are small. For the second scenario shown in Figure~\ref{fig:opt_scenario2}, the expected gross profits are overestimated for prices under 16 euros and underestimated for the remaining prices. The optimal price is still correctly determined, but the probability for the highest expected gross profit is also large for the price of 14.75 euros. Finally, Figure~\ref{fig:opt_scenario3} depicts the results for the third scenario. We can see that the estimated expected gross profits are too optimistic on average for all prices but the optimal price is again correctly estimated, however the price of 15.25 euros also has a high probability of yielding the highest expected gross profit. Overall, the differences are small compared to the true model, and the true expected gross profits still fall between the 95\% credible intervals for the price options considered in all scenarios and all prices.

\section{Discussion} \label{sec:discussion}

We proposed an approach for optimizing the price of a subscription-based service when real-life pricing experiments are unavailable. We described the customers' behavior in the form of a causal graph and identified the causal effect of the price by combining population statistics, purchase history, and conjoint data. The causal effect was estimated by Bayesian methods and as a result, we obtained a posterior distribution for the optimal price that maximizes the expected gross profit.

The strengths of the proposed approach are related to the use of causal inference to guide the fusion of data sources. Instead of mixing the historical observational data with experimental data from the conjoint studies in an ad~hoc fashion, we illustrated how such data sources can be combined in a theoretically sound fashion using causal inference. While we relied on simulated data due to the confidentiality of the real data, the causal graph used in the simulation and subsequent analysis were designed to emulate the assumed causal relations of the motivating real scenario. The approach worked well also in the real business application and was used to support pricing decisions.

We demonstrated that combining data sources is often necessary for price optimization. In our scenario, we were unable to estimate the model when the conjoint data were not available due to identifiability issues. Purchase history alone was insufficient for price optimization as the variation in the historical prices was low. However, the proposed approach is fairly robust to model misspecification in terms of price optimization.

The proposed approach requires a good understanding of the causal mechanisms of customer behavior. These mechanisms can be communicated in the form of a graph as illustrated in Figure~\ref{dag}. The most critical assumptions are those related to unobserved confounders because the causal effect of price on the purchase may not be identifiable in the presence of unobserved confounders.

Another important assumption is the stability of market conditions and customer behavior in time. Without this assumption, we could not use past data to make decisions about the future. In practice, this assumption means that the data used for the analysis must be recent, not decades old.

For the clarity of the presentation, the presented scenario included some simplifying assumptions that could be extended in real use cases. For instance, the number of background variables could be higher and customer-level data on service usage could be available. In some cases, it may be necessary to model the seasonal variation in the demand and the number of upcoming customers. The design of the conjoint study could also be more refined than the described setting. If a real-life pricing experiment is available, it should naturally be included as one of the data sources. The proposed approach can be easily adapted to these more complicated settings.

A possible direction for future research includes extension to multiple products. In the case of a subscription-based business model, there could be alternative subscription plans with different coverage, quality, and price. The model would be expanded to incorporate product-specific reference prices that are correlated at the customer level. Along similar lines, we could take into account the price changes implemented by the competitors. As the competitors are expected to react to price changes by the company in focus, dynamical modeling, and game theoretic considerations come into play.

Our work is one of the first applications of causality-driven data-fusion in the business context. We anticipate that the proposed methodology can be applied to various other causal inference and price optimization problems.

%

\section*{Declaration of interests}

ST and JK have applied the proposed method in consulting.


\acks{LV was supported by the Finnish Cultural Foundation, Central Finland Regional Fund and the Foundation for Economic Education. ST and JH were supported by the Research Council of Finland, grant number 331817.}


\newpage
\appendix
\section*{Appendix}



\begin{table}[!h]
\caption{Notation used in this paper. \label{tab:notation}}
\begin{center}
\spacingset{1.25}
\scriptsize
\begin{tabular}{ l l }
\textbf{DAG} & \\
$t$ & Time $t=1,2,3,\ldots$ \\
$B$ & Age $A$, Gender $G$, and Location $L$ of a customer\\
$Q$ & Customer's reference price related to the product\\
$C$ & Transportability node concerning $Y_t$ between real purchase and conjoint scenario \\
$S_t$ & Number of consecutive subscription periods of a customer until time $t$ \\
$X_t$ & Price of the product at time $t$\\
$Y_t$ & Customer's choice at time $t$\\
$H$ & Transportability node of differences between $B$ among subscribers and non-subscribers \\\hline
\textbf{Data} & \\
$\mathcal{D}^{pop}$ & Population\\
$\mathcal{D}^0$  & Customers at time $t$ having $S_t=0$ \\
$\mathcal{D}^1$ & Customers at time $t$ having $S_t>0$ \\
$\mathcal{D}$ & $\mathcal{D}^0 \cup \mathcal{D}^1$ i.e., Current customer data\\
$\mathcal{C}^0$ & Conjoint data of earlier subscribers\\
$\mathcal{C}^1$ & Conjoint data of current subscribers\\
$\mathcal{C}^2$ & Conjoint data of customers who have never subscribed\\
\hline
\textbf{Model} & \\
$\beta_0, \beta_{1,A}, \beta_{2,G}, \beta_{3,L}$ & Constant and regression coefficients of $B$ related to $Q$ \\
$u_{i}$ & Individual random effect in reference price of $i$th customer \\
$\tau$ & Standard deviation parameter of the random effect $u_i$  \\
$\pi_{i,t}$ & Purchase probability of the $i$th customer\\
$\alpha_1, \alpha_2, \alpha_3$ & Regression coefficients of ($Q + \kappa - X_t$), $S_t$, and $I(S_t=0$), respectively\\
$\kappa$ & Conjoint effect on $Y_t$ \\
$\mathbf{\Theta}$ & Common notation for all unknown parameters above \\\hline
\textbf{Optimization} & \\
$\mu^0_i(x,t)$ & Average purchase probability of the $i$th customer having $S_t=0$  \\
$\mu^1_i(x,t)$ & Average purchase probability of the $i$th customer having $S_t>0$  \\
$Y^0_{it}$ & Choice of the $i$th customer in data $\mathcal{D}^0_t$  \\
$Y^1_{it}$ & Choice of the $i$th customer in data $\mathcal{D}^1_t$  \\
$N^0(t)$ & Number of potential customers making a choice at time $t$\\
$N^1(t)$ & Number of current customers making a choice at time $t$\\
$f$ & Expected gross profit of the product\\
$\mathcal{V}$ & Variable costs of the product \\
\end{tabular}
\end{center}
\end{table}

\begin{table}[ht]
\caption{True values, posterior means, posterior standard deviations, credible intervals, $\hat{R}$ estimates, and effective sample sizes of the model parameters obtained from the MCMC simulation of the model using purchase history and conjoint data. Individual-specific random effects are not shown.}
\centering
\spacingset{1.25}
\footnotesize
\begin{tabular}{lrrrrrrrr}
Parameter & True & Mean & SD & 2.5\% & 97.5\% & $\hat{R}$ & Bulk-ESS & Tail-ESS \\
  \hline
$\beta_0$               &  $2.8000$ &  $2.8072$ & $0.0072$ &  $2.7955$ &  $2.8188$ & $1.0006$ & $4964$ & $4594$ \\
$\beta_{1, A=18-30}$    &  $0.0000$ &           &          &           &           &          &        & \\
$\beta_{1, A=31-45}$    & $-0.0150$ & $-0.0139$ & $0.0050$ & $-0.0223$ & $-0.0058$ & $0.9994$ & $4962$ & $4625$ \\
$\beta_{1, A=46-60}$    & $-0.0300$ & $-0.0263$ & $0.0051$ & $-0.0347$ & $-0.0180$ & $1.0004$ & $4734$ & $4688$ \\
$\beta_{1, A=61-75}$    & $-0.0450$ & $-0.0444$ & $0.0054$ & $-0.0534$ & $-0.0356$ & $1.0005$ & $4758$ & $4402$ \\
$\beta_{2, G="male"}$   &  $0.0000$ &           &          &           &           &          &        & \\
$\beta_{2, G="female"}$ &  $0.0100$ &  $0.0067$ & $0.0036$ &  $0.0008$ &  $0.0125$ & $1.0013$ & $4748$ & $4862$ \\
$\beta_{3, L="urban"}$  &  $0.0000$ &           &          &           &           &          &        & \\
$\beta_{3, L="rural"}$  & $-0.0200$ & $-0.0160$ & $0.0042$ & $-0.0227$ & $-0.0091$ & $1.0016$ & $4733$ & $4759$ \\
$\tau$                  &  $0.1000$ &  $0.0872$ & $0.0077$ &  $0.0743$ &  $0.1000$ & $0.9995$ & $4319$ & $4624$ \\
$\alpha_1$              &  $0.3500$ &  $0.3458$ & $0.0108$ &  $0.3278$ &  $0.3636$ & $1.0005$ & $4763$ & $4782$ \\
$\alpha_2$              &  $0.4500$ &  $0.4751$ & $0.0508$ &  $0.3908$ &  $0.5580$ & $1.0000$ & $4490$ & $4308$ \\
$\alpha_3$              & $-0.3000$ & $-0.3303$ & $0.0387$ & $-0.3947$ & $-0.2666$ & $1.0009$ & $4852$ & $4663$ \\
$\kappa$                &  $0.7500$ &  $0.8277$ & $0.1058$ &  $0.6553$ &  $1.0020$ & $1.0004$ & $4603$ & $4779$ \\
   \hline
\end{tabular}
\label{param_estimates}
\end{table}

\clearpage


\end{document}